\documentclass[aapm,reprint,mph,longbibliography]{revtex4-1}

\usepackage{graphicx}
\graphicspath{ {figures/} }
\newcommand{\virtual}{\emph{virtual}\ }

\usepackage{color}

\usepackage[separate-uncertainty = true,multi-part-units=single]{siunitx}

\usepackage{hyperref}

\usepackage{amsmath}

\usepackage{epstopdf}

\begin{document}

\title{Technical Note: Towards Virtual Monitors for Image Guided Interventions\\Real-time Streaming to Optical See-Through Head-Mounted Displays} 



\author{Long Qian, Mathias Unberath, Kevin Yu, Nassir Navab}
\affiliation{Computer Aided Medical Procedures, Johns Hopkins University}

\author{Bernhard Fuerst}
\affiliation{Computer Aided Medical Procedures, Johns Hopkins University\\Bernhard Fuerst is now with Verb Surgical.}

\author{Alex Johnson, Greg Osgood}
\affiliation{Department of Orthopaedic Surgery, Johns Hopkins University}

\date{\today}

\begin{abstract}
\noindent
\textbf{Purpose:}
Image guidance is crucial for the success of many interventions. Images are displayed on designated monitors that cannot be positioned optimally due to sterility and spatial constraints. This indirect visualization causes potential occlusion, hinders hand-eye coordination, leads to increased procedure duration and surgeon load.\\
\textbf{Methods:} 
We propose a \virtual monitor system that displays medical images in a mixed reality visualization using optical see-through head-mounted displays. The system streams high-resolution medical images from any modality to the head-mounted display in real-time that are blended with the surgical site. It allows for mixed reality visualization of images in head-, world-, or body-anchored mode and can thus be adapted to specific procedural needs.\\
\textbf{Results:}
For typical image sizes, the proposed system exhibits an average end-to-end delay and refresh rate of \SI{214\pm30}{\milli\second} and \SI{41.4\pm32.0}{\hertz}, respectively.\\
\textbf{Conclusions:}
The proposed \virtual monitor system is capable of real-time mixed reality visualization of medical images. In future, we seek to conduct first pre-clinical studies to quantitatively assess the impact of the system on standard image guided procedures.
\end{abstract}


\maketitle 

\section{Introduction}
Every day, countless image guided interventions are conducted by a diverse set of clinicians across many disciplines. From procedures performed by ultrasound technicians \cite{bajura1992merging,rumack2005diagnostic,peterson2017introduction}, to orthopaedic surgeons \cite{strobl2014technical,de2017c}, to interventional radiologists \cite{tacher2013image,mason2014accuracy}, one aspect unites them all: the viewing of medical images on conventional monitors \cite{westwood2005mini,yaniv2006image,hallifax2014physician}.\\
In many of the aforementioned interventional scenarios, real time images are acquired to guide the
procedure. However, these images can only be viewed on designated wall mounted monitors. The ability
to position these displays is limited due to sterility, flexibility as they are bound to mounts, and the  spatial constraints of the room such as the operating team and equipment. Consequently, images designated for procedural guidance cannot be displayed “in-line” with the operative field \cite{cardin2005method,westwood2005mini,yaniv2006image}. This indirect visualization with images visually off-axis from the intervention site has been shown to create a disconnect between the visuo-motor transformation hindering hand-eye coordination \cite{wentink2001eye}. Situations that allow for the viewing of one’s hands and the guiding image simultaneously with an “in-line” view helps to solve this problem \cite{hanna1998task,erfanian2003line,westwood2005mini,chimenti2015google}. To alleviate this problem, previous approaches placed miniature LCD displays close to the intervention site \cite{cardin2005method,westwood2005mini} or displayed images via Google Glass \cite{chimenti2015google,yoon2016technical}. Unfortunately in all these cases, the small size and poor resolution of these displays limits the conveyable information impeding standalone use and, hence, clinical relevance \cite{yaniv2006image}.\\
Recent advances in optical see-through head-mounted display (OST-HMD) technology enable high resolution, binocular displays directly in the field of vision of the user without obstructing the rest of the visual scene \cite{keller2008head,iqbal2016review}. Coupled with medical imaging, this technology may provide \virtual monitors that can be positioned close to the intervention site and are large enough to convey all required information. This technology has the potential to overcome aforementioned drawbacks. Our hypothesis is that the use of \virtual displays based on OST-HMDs that enable “in-line” image guidance will allow clinicians to perform procedures with higher efficiency and with improved ergonomics over conventional monitors.\\
Within this technical note, we describe a setup for real-time streaming of high-resolution medical images to the Microsoft HoloLens \cite{hololens} that enables head-, world-, and body-anchored \virtual monitors within the operating suite.

\section{Material and Methods}

\subsection{System Overview}

\begin{figure}[t]
    \centering
    \includegraphics[width=0.95\linewidth]{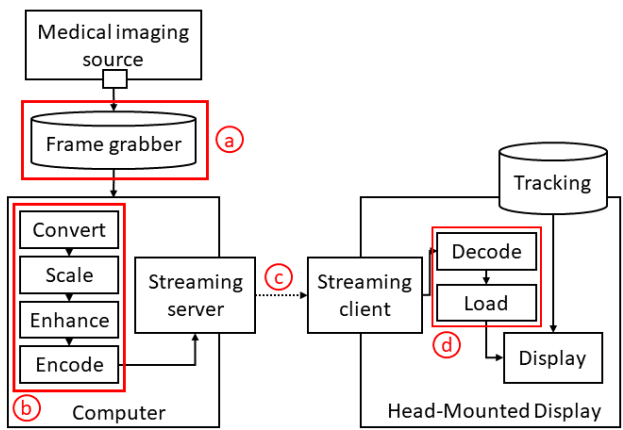}
    \caption{Schematic overview of the system components and key functionality. Colored annotations highlight functionality modules that are potential bottlenecks for real-time streaming. Their individual performance is evaluated separately in \autoref{sec:validation}.}
    \label{fig:pipeline}
\end{figure}

The concept of a virtual monitor for image-guided procedures can be realized via real-time streaming of the intra-procedurally acquired medical images or image sequences, i.\,e. video, to an optical see-through head-mounted display. The OST-HMD then visualizes the images, blending them with the reality perceived by the wearer. In the case presented here, we assume that the HMD is equipped with a tracking module. Then, medical images can be displayed in different modes allowing for different mixed reality experiences. A more detailed description of this circumstance is given in \autoref{sec:mixedReality}.\\
\autoref{fig:pipeline} demonstrates the components and functionalities of the proposed system. Components are introduced in \autoref{sec:materials} while functionalities are described in detail in \autoref{sec:methods}.

\subsection{Material}
\label{sec:materials}

\subsubsection{Medical Imaging Source}
Medical imaging sources provide input to the proposed real-time streaming pipeline. Potential medical imaging sources include 2D imaging modalities such as diagnostic X-ray fluoroscopy systems, interventional C-arm cone-beam scanners, and Ultrasound systems but naturally extends to 3D image sources including computed tomography, cone-beam computed tomography, and magnetic resonance imaging.\\ 
Traditionally, medical images are transferred within a vendor-specific framework inside the operating room. We use a video output port provided by the manufacturer to tap the medical imaging data after internal pre-processing that is simultaneously supplied to the traditional radiology monitors.

\subsubsection{Frame Grabber}

The frame grabber is hardware that is connected to a video output port of any imaging source (in this case a medical imaging modality) and has access to the imaging data. In setups where the medical imaging source provides an interface for direct access to the data, the frame grabber is not a necessary component. However, use of a frame grabber has the additional benefit that it effectively decouples medical image generation and internal pre-preprocessing and the proposed streaming pipeline into two separate closed loops, such that the traditional imaging pipeline in the operating room remains unaffected. 

\subsubsection{Image Processing Framework}

The image processing framework is responsible for converting, scaling, enhancing, and encoding the image at runtime. Memory-inefficient pixel formats can be converted to more efficient pixel format that allow for faster processing and transfer, e.\,g., a conversion from RGBA32 to YUV2, or to gray-scale. Scaling refers to the manipulation of the pixel size of the image and constitutes a trade-off between processing load and image quality. Enhancing is an optional step in the image processing pipeline. Well-known representatives of image processing filters are, e.\,g., contrast enhancement or denoising \cite{sonka2014image}, that can be employed to further improve the perception and readability of the visualized medical images. Encoding describes the process of compressing images or fragmenting the data into smaller packets to enable efficient transfer or storage. Common encoders include, among others, Motion-JPEG \cite{mohr2009mjpeg}, H264 \cite{marpe2006h264}. Motion-JPEG is used in our setup.

\subsubsection{Data Transfer Network}

Data packets are transferred from the image processing framework to head-mounted display via a data transfer network. Depending on the specifications of the particular image processing framework and HMD device, the data transfer may happen locally, via cable, or via wireless router. For the setup described here we assume use of an untethered device. The image processing framework is realized on a stationary computer and, consequently, a wireless router (NETGEAR Nighthawk R6700 \cite{netgearRouter}) is used for communication and data transfer. TCP/IP \cite{parziale2006tcp} acts as the communication protocol. 

\subsubsection{Head-Mounted Display}

The HMD receives the data packets from the data transfer network, decodes the data packets into images, and loads the decoded images into the rendering engine. Finally, it visualizes the sequence of images in a mixed reality environment with the help of tracking module. Within our experiments the Microsoft HoloLens \cite{hololens,kress2017hololens} is used as the mixed reality headset. An exemplary image of this OST-HMD that was released in March 2016 \cite{kress2017hololens} is provided in \autoref{fig:transform}.

\subsection{Methods}
\label{sec:methods}

\subsubsection{Tracking and Localization}
\label{sec:tracking}

\begin{figure}[t]
    \centering
    \includegraphics[width=0.9\linewidth]{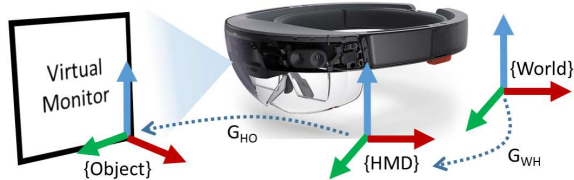}
    \caption{Relevant transformations for visualization in mixed reality environments. The figure shows a Microsoft HoloLens, the image of which is taken from \cite{hololensImage}.}
    \label{fig:transform}
\end{figure}

Tracking and localization is the enabling mechanism for different mixed reality rendering effects. The virtual monitor effect requires the HMD to maintain knowledge about its position and orientation in the operating room, involving both hardware sensors and software algorithms. Common methods of tracking can be categorized into outside-in methods, such as external optical tracking \cite{khadem2000comparative}, and inside-out approaches, e.\,g., simultaneous localization and mapping (SLAM) \cite{durrant2006slam}, that are employed in the scenario described here \cite{kress2017hololens}. The transformations between the coordinate systems of world, HMD and visualized object (here the \virtual monitor) are demonstrated in \autoref{fig:transform}. Particularly, $G_{WH}$ is the transformation from the world to the HMD coordinate system and is computed from the tracking module. $G_{HO}$ describes the mapping from the HMD to the virtual object coordinate that is rendered on the HMD; it is controlled by the rendering algorithm. With Microsoft HoloLens, $G_{WH}$ is computed from SLAM algorithms \cite{quian2017calib} and is available in real-time.\\
In the preceding description we have omitted that aforementioned transformations may not be constant over time. In fact, tracking of the relative movement of these coordinate systems is a key asset of the proposed pipeline, the reason of which is detailed below.

\subsubsection{Mixed Reality Visualization}
\label{sec:mixedReality}

Aided by tracking and localization methods, the user is free to chose between three kinds of mixed reality visualization modes that are presented in the following sections. This flexibility in mixed reality experience is one of the major advancement of the proposed compared to earlier systems \cite{chimenti2015google,hannaford2017heads}.

\paragraph{Head-Anchored}

In the head-anchored visualization, the rendered object is placed at a fixed pose relative to the user. This means that $G_{HO}$ remains constant. In consequence, medical images are visualized in a head-up display manner. Researchers have exploited the benefits of head-anchored visualization in \cite{vorraber2014medical,chimenti2015google,hannaford2017heads}. Head-anchored visualization is powerful as it makes full use of the HMD in terms of visualization of the content. However, for cases where the surgeon does not want the medical images occluding potentially crucial areas of the operating field, head-anchored visualization is a distraction.

\paragraph{World-Anchored}

The world-anchored visualization is closest to current clinical practice. It creates a \virtual monitor effect, where the user is able to see a medical imaging display as if it was presented on a traditional monitor. The 6 degree-of-freedom pose of the virtual object is invariant in the world coordinate system, i.\,e, 
\begin{align}
  G_{WO}(t) &= G_{HO}(t)\cdot G_{WH}(t) = \text{const.} \nonumber \\
  G_{HO}(t) &= G_{WO}(t)\cdot G_{WH}(t)^{-1}\,,
\end{align}
where $t$ denotes the current time point as motivated in \autoref{sec:tracking}.\\
The rendering framework needs to incorporate real-time tracking results, and adjusts the pose of the virtual object accordingly. World-anchored visualization within the medical context has been studied in \cite{chen2015development,wang2016precision}. 
World-anchored visualization is intuitive as it resembles the traditional monitor, and gives more control to the user in terms of the display configuration, e.\,g., with respect to location, orientation, and brightness. 

\paragraph{Body-Anchored}

Body-anchored display is a concept that blends both head-anchored and world-anchored display. When the extent of the user's motion is large, the rendered virtual object follows the user's motion similar to head-anchored visualization. While the virtual object remains in the field-of-view of the user at all times, it is not necessarily always at the same pose as it would be in a head-anchored display. On the other hand, when the motion is small, which often happens when the user is slightly adjusting the viewing perspective to better perceive the virtual object, the virtual object remains fixed in the world space as a world-anchored display.

\subsection{Experimental Validation}
\label{sec:validation}

We quantitatively assess the performance of the key components of the proposed pipeline individually which then enables the computation of the total end-to-end performance. All components evaluated in this manner are highlighted by red boxes in \autoref{fig:pipeline}. This procedure allows for the identification of potential bottlenecks for system performance. All described measurements were repeated sufficiently many times ($>30$) to ensure a representative set of samples.

\subsubsection{Frame Grabber}
We retrieve the delay of the frame grabber (component (a) in \autoref{fig:pipeline}) via a virtual stopwatch with a resolution of \SI{15}{\milli\second} and display the current time on the screen. The screen is then captured by the frame grabber, streamed to the same computer, and visualized in split screen. This technique allows for the computation of delay as we simultaneously display the current system and the captured video frame time.\\ 
The refresh rate of the frame grabber depends on the screen's resolution and computation power of the receiving computer. In our tests, we used an Epipahan DVI2USB 3.0 and an Alienware 15 R3 Laptop with Intel(R) Core i7-7700HQ equipped with a NVIDIA GeForce GTX 1070 operated at \SI{1920x1080}{} pixels.

\subsubsection{Image Processing}
For the image processing pipeline, denoted by (b) in \autoref{fig:pipeline}, we assess the time between the moment the frame is available to the point when the frame is ready to be sent via the data transfer network. In our experiment, we use an incoming frame size of \SI{1920x1080}{} pixels in RGB24 that is then down-sampled (scaled) to \SI{800x450}{} pixels still in RGB24 format.

\subsubsection{Data Transfer Network}
Under normal circumstances, latency refers to the time that elapses between sending the frame from the host to receiving the same frame on the device, i.\,e. the HoloLens. Measuring latency in this setup, however, is complicated due to imperfect synchronization of host and device. In order to avoid inaccuracies due to synchronization, we rely on the HoloLens Emulator \cite{hololensEmulator}. Both, emulator and server, are running on the same machine thus sharing system time. In \autoref{fig:pipeline}, this delay is indicated as (c). 

\subsubsection{Head-Mounted Display}
We measure the time on the device between receiving an image and loading the frame into the texture that is used for visualization, a delay that is denoted by (d) in the system overview shown in \autoref{fig:pipeline}. Moreover, We assess the rate of frame reception on the device and evaluate the texture loading time and the rendering refresh rate for multiple image sizes.\\
A more thorough evaluation of the display and tracking capabilities of a candidate OST-HMD is meaningful and important, particularly when considering applications in the medical context. This evaluation, however, is beyond the scope of the presented technical note. We would like to refer to a previous in-house study that compares multiple devices with respect to aforementioned criteria \cite{qian2017comparison}.

\section{Results}
\label{sec:results}
We state the average delay $\mu$ and the respective standard deviation $\sigma$ for every of the aforementioned measurements in \autoref{tab:delays}. In summary, we found an average end-to-end transmission time of \SI{214}{\milli\second} (\SI{30}{\milli\second}). The grabbing refresh rate was found to be a constant \SI{36}{\hertz}. Further, the number of received frames depending on the image size is stated in \autoref{tab:framesReceived}. 
Finally, results for the texture loading time and refresh rate as a function of input image size are provided in \autoref{tab:fps}. Very high refresh rates above \SI{30}{\hertz} are achieved for image sizes smaller \SI{1000x1000}{} pixels in RGB24.

\begin{table}[tb]
  \centering
  \caption{Average delay $\mu$ and standard deviation $\sigma$ for the key components of the proposed network stated as $\mu\ (\sigma)$. All values are stated in \SI{}{\milli\second}.}
  \label{tab:delays}
  \begin{tabular}{| c | c | c | c |}
  	\hline
    (a) Grabbing & (b) Processing & (c) Transfer & (d) HMD\\ \hline\hline
    106 (29) & 100 (7) & 6.03 (2.28) & 2.12 (3.71) \\
    \hline
  \end{tabular}
\end{table}

\begin{table}[tb]
  \centering
  \caption{Average number $\mu$ and standard deviation $\sigma$ of received frames on the HMD device. All values stated in \SI{}{\hertz}.}
  \label{tab:framesReceived}
  \begin{tabular}{| c | c | c | c | c |}
  	\hline
    \SI{620x480}{} & \SI{800x450}{} & \SI{1000x1000}{} & \SI{1600x900}{} & \SI{1920x1080}{} \\ \hline\hline
    62.2 (44.6) & 41.4 (32.0) & 31.8 (23.7) & 27.6 (19.4) & 31.7 (26.2) \\
    \hline
  \end{tabular}
\end{table}

\begin{table}[tb]
  \caption{Overall rendering refresh rate in \SI{}{\hertz} as a function of input image size. Images are RGB24.}
  \label{tab:fps}
  \begin{tabular}{|l||c|c|}
    \hline
    Image size& (d) HMD $\mu\ (\sigma)$ in \SI{}{\milli\second} & Rendering $\mu$ $(\sigma) in \SI{}{\hertz}$\\
    \hline
    \hline
    \SI{620x480}{}		&3.33 (8.83)	&46.4 (6.7)\\
    \SI{800x450}{}		&3.92 (10.47)	&47.1 (6.5)\\
    \SI{1000x1000}{}	&16.9 (34.2)	&32.1 (9.2)\\
    \SI{1600x900}{}		&25.7 (49.3)	&22.5 (9.0)\\
    \SI{1920x1080}{}	&61.3 (81.7)	&16.4 (6.5)\\
    \hline
  \end{tabular}
\end{table}

\section{Discussion}
From our system evaluation presented in \autoref{sec:validation} and \autoref{sec:results}, we computed an average end-to-end delay of \SI{214}{\milli\second} (\SI{30}{\milli\second}) that may be sufficient for procedures that do not require very fast tool motion.\\
We have observed very high effective end-to-end frame rates of above \SI{30}{\hertz} for image sizes larger than \SI{1000x1000}{} pixels in RGB24. Our findings suggest that, despite the large standard deviation of frame reception rate stated in \autoref{tab:framesReceived}, the effective frame rate of the proposed \virtual monitor system is limited by the texture rendering of the device.\\ 
Effective frame rates of \SI{30}{\hertz} with an average delay of \SI{214}{\milli\second} may not yet be convincing for optical video guided interventions using, e.\,g., endoscopy. However, we believe that the proposed system is fit for deployment in first pre-clinical studies in X-ray guided procedures. X-ray fluoroscopy is usually operated on lower frame rates (\SIrange{7.5}{30}{\hertz}) to reduce exposure for both the patient and the surgeon \cite{mahesh2001fluoroscopy}.\\ 
From the detailed analysis shown in \autoref{tab:delays} we identified frame grabbing and image processing as the major bottlenecks with respect to delay, taking on average more than \SI{100}{\milli\second} each. The frame grabber comes with its proprietary driver and software and does not easily allow tweaks for performance. The image processing pipeline, however, potentially allows for modifications in future work that could target more efficient scaling and compression of the images.\\
The high standard deviation of the frame reception on the HMD device suggests bunching of incoming images. This behavior is not desirable, as it implies a varying frame rate and, thus, credibility of the displayed images at different time points during an image guided intervention. Future work will hence investigate possibilities of homogenizing the data transfer over time.

\section{Conclusion}
We have presented and evaluated an system for real-time visualization of medical images in a mixed reality surgical environment using an optical see-through head-mounted display. The system comprises of a frame grabber, an image processing unit, a data transfer network, and a head-mounted display. It accepts images of any conventional medical imaging modality. The current setup allows for real-time mixed reality visualization of medical images in head-, world-, and body-anchored display with an average delay of \SI{214\pm30}{\milli\second} and a refresh rate of around \SI{41.4\pm32.0}{\hertz} for typical image sizes.\\
Given the promising results, we are excited to assess the impact of the system on the surgical work flow that we seek to study in first pre-clinical tests. Future improvements to the system will investigate means of stabilizing the refresh rate of the \virtual monitor.


\section*{Conflicts of interest}
The authors have no relevant conflicts of interest to disclose.

%

\end{document}